\documentclass[11pt]{article}
\usepackage{aaspp4}

\received{05/31/99}
\accepted{09/14/99}

\slugcomment{Astrophysical Journal}
\lefthead{BAUER \& SARAZIN}
\righthead{X-RAY PROPERTIES OF ABELL 644}

\begin{document}

\title{X-ray Properties of the Abell 644 Cluster of Galaxies}

\author{Franz Bauer and Craig L. Sarazin}

\affil{Department of Astronomy, University of Virginia,
P.O. Box 3818, Charlottesville, VA 22903-0818;
feb4q@virginia.edu, cls7i@virginia.edu}

\begin{abstract}
We use new {\it ASCA} observations and archival {\it ROSAT} Position
Sensitive Proportional Counter (PSPC) data to determine the X-ray
spectral properties of the intracluster gas in Abell 644.  From the
overall spectrum, we determine the average gas temperature to be
$8.64^{+0.67}_{-0.56}$ keV, and an abundance of $0.32\pm0.04$
$Z_{\odot}$.  The global {\it ASCA} and {\it ROSAT} spectra imply a
cooling rate of $214^{+100}_{-91}$ $M_{\odot}$ yr$^{-1}$.
The PSPC X-ray
surface brightness profile and the {\it ASCA} data suggest a somewhat
higher cooling rate.
We determine the gravitational mass and gas mass as a function of
radius.  The total gravitating mass within 1.2 Mpc is
$6.2\times10^{14}$ $M_{\odot}$, of which 20$\%$ is in the form of hot
gas.  There is a region of elevated temperature 1.5-5$^{\prime}$ to
the west of the cluster center.  The south-southwest region of the
cluster also shows excess emission in the {\it ROSAT} PSPC X-ray
image, aligned with the major axis of the optical cD galaxy in the
center of the cluster.
We argue that the cluster is undergoing or has
recently undergone a minor merger.
The combination of a fairly strong
cooling flow and evidence for a merger make this cluster an
interesting case to test the disruption of cooling flow in mergers.
\end{abstract}

\keywords{cooling flows ---
dark matter ---
galaxies: clusters: general ---
galaxies: clusters: individual (Abell~644) ---
intergalactic medium ---
X-rays: galaxies
}

\clearpage

\section{Introduction} \label{sec:intro}

Cooling flows occur in clusters of galaxies where the age of the
cluster has exceeded the cooling time scale of intracluster gas.  In
the more luminous clusters, this cooling time is comparable to the
Hubble time (and thus cluster ages). Thus we should expect to see these
flows, provided they have remained in hydrostatic equilibrium since 
formation. This result has been validated by studies with {\it Einstein}, {\it
Ginga}, and {\it ROSAT} which indicate that cooling flows are common and
long-lived phenomenon among nearby clusters
(Edge et al.\ 1992;
Peres et al.\ 1998). 

Abell 644 (A644), with a bolometric luminosity exceeding $10^{45}$
ergs s$^{-1}$ and lying at a redshift of $z = 0.0704$, is one of the
brighter clusters in the local universe.
However, A644 is only a
richness class 0 cluster (Abell, Corwin, \& Olowin 1989), and thus has
a higher than average X-ray luminosity for its richness class
(typically $L_X = 10^{42-44}$ ergs s$^{-1}$;
Abramopoulos \& Ku 1983;
Briel \& Henry 1993). This cluster has been 
previously observed in X-rays with the {\it Einstein}, {\it Ginga},
and {\it ROSAT} observatories.  The {\it Einstein} observations of
A644 implied a significant cooling flow, with an accretion rate of 326
$M_{\odot}$ yr$^{-1}$ (Edge et al.\ 1992), and an average gas
temperature of 7.2 keV (David et al.\ 1993). More recent analysis of
the {\it ROSAT} PSPC and HRI data imply slightly lower values of
$\sim$200 $M_{\odot}$ yr$^{-1}$ and 6.6 keV
(Peres et al.\ 1998;
White, Jones, \& Forman 1997).

We present new {\it ASCA} X-ray spectra observations of A644
(\S~\ref{sec:observ}). The use of {\it ASCA} spectra to determine the
spatial variation of the spectral properties of the X-ray emitting gas
in A644 requires that the effects of the energy-dependent point-spread
function (PSF) be corrected. We also analyze archival {\it ROSAT} PSPC
observations of the X-ray image and spectrum of A644. These spectra
allow us to constrain global parameters (\S~\ref{sec:global}) as well
as temperature variations and abundance gradients in the intracluster
gas (\S~\ref{sec:annular}).  We use the
{\it ASCA} derived temperature gradients and {\it ROSAT} derived X-ray
surface brightness to determine the radial variation of the total
gravitational mass and gas mass in the cluster (\S~\ref{sec:mass}). We
test the hydrostatic assumption by searching for evidence of
asymmetries in the temperature distribution (\S~\ref{sec:sections}),
which would indicate recent dynamical activity, such as a subcluster
merger. We also estimate the cooling rate
(\S~\ref{sec:mdot}). We summarize our results and discuss their
implications in \S~\ref{sec:conclusion}.
Throughout this paper, we assume a Hubble constant $H_{o} = 50$ km
s$^{-1}$ Mpc$^{-1}$ and a cosmological deceleration parameter of
$q_{o} = 0.5$. At a redshift of $z = 0.0704$,
the angular diameter distance to the cluster is 375 Mpc,
and $1\arcmin$ corresponds to 109 kpc.
Unless otherwise stated, all errors are 90$\%$ confidence intervals for one
parameter of interest.

\section{Observations and Data Selection} \label{sec:observ}

A644 was observed with {\it ASCA} on 1995 April 18-20 for a total of
66.3 ksec with the two Gas Imaging Spectrometer (GIS) detectors and
58.3 ksec with the two Solid-State Imaging Spectrometer (SIS)
detectors.  The SIS observations were taken in 1~CCD faint mode, with
the cluster center placed near the center of chip 1 of SIS0 and chip 3
of SIS1.  We applied standard cleaning procedures (Day et al.\ 1995),
and selected data with a minimum cut off rigidity, elevation angle,
and angle from the sunlit Earth of 6 GeV/c, 10$^{\circ}$, and
25$^{\circ}$, respectively.  The screened exposure times were 49.5
ksec for the SIS detectors and 57.6 ksec for the GIS detectors.

We also retrieved the {\it ROSAT} PSPC observation of A644 from the
archive (RP800379N00; PI: B\"ohringer).  A644 was observed for 10.2
ksec on 1993 April 28-29.  The PSPC data were filtered for periods of
high background and other problems, and corrected for non-X-ray
background, vignetting, and exposure using the computer programs
developed by Snowden (Plucinsky et al.\ 1993; Snowden 1995). After
filtering, the live exposure was 8.3 ksec.  The average Master Veto
Rate for the filtered data was 83.7.

The analysis of the global cluster spectrum and the {\it ROSAT} PSPC
spectrum were done primarily with the XSPEC package (version 10.0),
and the analysis of the PSPC image was done using the PROS package
within IRAF.  The deconvolution of the spatial variation of the
spectrum from the energy-dependent Point Spread Function (PSF) of {\it
ASCA} was performed using the algorithms of Markevitch (Markevitch
1996; Markevitch et al.\ 1998).

\section{Global X-Ray Properties} \label{sec:global}

\placefigure{fig:contour}

Figure~\ref{fig:contour} shows a contour plot of the {\it ROSAT} PSPC
X-ray image, superposed on a optical image from the Digital Sky Survey
(Lasker et al.\ 1990).  The image was corrected for background,
vignetting, and exposure, and was smoothed with an adaptive kernel
which gave a minimum signal to noise ratio of 5 per smoothing beam
(Huang \& Sarazin 1996).  The cluster is fairly regular, but shows an
asymmetric extension to the SSW in both the X-ray and the optical
images; a similar extension was seen in the {\it Einstein} image (Mohr, 
et al.\ 1995). The cluster emission is strongly peaked on the position 
of the cD galaxy and the elongated X-ray structure is aligned with the cD 
major axis, an effect often seen in richer clusters.

The global X-ray spectrum of the cluster was determined by accumulating
the spectra from the {\it ROSAT} PSPC and the {\it ASCA} GIS
instruments in a circular region with a radius of 10$\arcmin$.  Beyond
this radius, the X-ray surface brightness rapidly diminishes.  We
excluded the SIS instruments because the field of view (FOV) is only
11$^{\prime}$ square in 1~CCD mode, and doesn't cover the entire
region used for the global cluster spectrum. However, including the 
SIS instruments does not significantly change the global results.  
The background for the {\it ASCA} observation was obtained from long
exposure background fields at similar galactic latitude, and was
extracted and cleaned identically to the data.  The energy ranges
adopted for the GIS was 0.6-10.0 keV.  The background for the {\it
ROSAT} PSPC spectrum was extracted from the same observation, using an
annulus of 34-44$^{\prime}$ and masking out any point sources.  The
X-ray background was corrected for vignetting.  Only energies from
0.24--2.5 keV were used for the fit to PSPC data.  All of the channels
used in the spectral fitting had more than 20 counts, so that the
$\chi^2$ distribution should be applicable.

Initially we fit both the {\it ASCA} and {\it ROSAT} data to a single
temperature Raymond-Smith optically thin thermal emission model
(Raymond \& Smith 1977), with the photoelectric absorption component
(Morrison \& McCammon 1984) fixed at the Galactic neutral hydrogen
column density, $N^{gal}_{H} = 8.45 \times 10^{20}$ cm$^{-2}$ (Stark
et al.\ 1992).  We will refer to this model as the Single Temperature
Model.  A MEKAL Mewe-Kaastra thin-thermal plasma model (Mewe,
Gronenschild, \& van den Oord 1985; Kaastra 1992) was fit as an
alternative, yielding nearly identical results.  At first, we
determined the temperature and overall heavy element abundance by
fitting the {\it ASCA} and {\it ROSAT} spectra separately.  The
relative abundance ratios of heavy elements were fixed at the solar
values given by Anders \& Grevesse (1989).

\placetable{tab:global}

The result of the single temperature fit to the {\it ASCA} GIS
detectors is shown in the first row of Table~\ref{tab:global} in the
columns labeled ``Single Temperature Model.''
This is an acceptable
fit, and gives reasonably well-determined values for the temperature
and heavy element abundance.  The result of the single temperature fit
to the {\it ROSAT} PSPC spectrum is shown in the second row of
Table~\ref{tab:global}.  This fit is poorly constrained by the data.
In addition, it gives a value for the abundance which is inconsistent
with the value given by {\it ASCA}, and much higher than values found
for similar clusters.  The third row of Table~\ref{tab:global} gives
the result of a joint {\it ASCA} and {\it ROSAT} PSPC fit and is
strongly skewed towards the {\it ASCA} results because the {\it ASCA}
fit was more strongly constrained than that of the {\it ROSAT} data.
Given the nearly disjoint energy ranges of {\it ASCA} and {\it
ROSAT}, the conflict in the two spectral fits and the poor fit to the
{\it ROSAT} spectrum suggest that the Single Temperature Model lacks a
soft X-ray emission or absorption component which is present in the
cluster.  We consider two possibilities; an excess soft X-ray
absorption or a cooling flow emission component.

First, we allowed the soft X-ray absorption to vary.  This improved
the fit somewhat, particularly for the {\it ROSAT} spectrum.  In
Table~\ref{tab:global}, we show the results of including an excess
absorber with a column density of $\Delta N_H$ and a covering factor
of unity at the redshift of the cluster (Excess Absorption Model).
The combined fit of the data give an overall temperature of 7.59 keV,
an overall heavy element abundance of 0.30 relative to solar, and an
intrinsic absorption column of $\Delta N_H < 1.26 \times 10^{20}$
cm$^{2}$.  The $\chi^2$ for the fit with excess absorption is improved
by 2.4 for one extra fitting parameter, which is significant at the 
85$\%$ level for the f-test. However, the excess column is only about 
15$\%$ of the nominal Galactic column of $N^{gal}_{H} = 8.45 \times 
10^{20}$cm$^{-2}$ (Stark et al.\ 1992), and thus could easily result 
from interpolating the H~I data column to the position of the cluster.
Moreover, for Galactic lines--of--sight with columns $N^{gal}_{H} \ga
5 \times 10^{20}$ cm$^{-2}$ such as A644, there is a significant
molecular component which increases the X-ray absorbing columns above
those of H~I (e.g., Arabadjis \& Bregman 1999).  Thus, it seems likely
that the small excess absorption required to fit the X-ray spectrum is
Galactic in origin. However, the fact that the best-fit PSPC
temperature is still completely inconsistent with the ASCA temperature
indicates that a soft emission component is still missing.

Second, we added a cooling flow spectrum to the model, assuming the
gas cooled isobarically subject to its own radiation from the ambient
temperature in the central region to a very low temperature.  The
cooling gas had the same abundances as the ambient gas. The results
are listed in the third column of Table~\ref{tab:global} (Cooling Flow
Model). The global fit to {\it ASCA} GIS is significantly better. The
{\it ROSAT} temperature, while barely bounded, is now consistent with
{\it ASCA} and the abundances match. The fit to the {\it ROSAT} PSPC
spectra is almost unchanged, indicating that these spectra do not
require a cooling flow component. We derive a cooling rate of
$214^{+100}_{-91}$ $M_\odot$ yr$^{-1}$.  In the combined fit,
the addition of the cooling flow (with one more fitting parameter)
decreases $\chi^2$ by 16.9, which is a significant decrease at the
99.9$\%$ confidence level for the f-test.  Thus, there is
evidence for a cooling flow, at roughly the same rate as previous
estimates.

The flux and luminosity, as derived from {\it ASCA} in the energy 
range of 2.0-10 keV are $5.45 \times 10^{-11}$ ergs cm$^{-2}$ s$^{-1}$
and $1.24 \times 10^{45}$ ergs s$^{-1}$, respectively, for a circular 
region with a radius of 10$\arcmin$ centered on A644.  Previous
measurements with {\it Einstein} MPC gave a cluster temperature of
$7.2^{+1.8}_{-1.2}$ keV, a flux and luminosity of $4.13 \times
10^{-11}$ ergs cm$^{-2}$ s$^{-1}$, and $9.55 \times 10^{44}$ ergs
s$^{-1}$ using the 2-10 keV energy band and a circular aperture of 
$45\arcmin$ (David et al.\ 1993).  The
previous temperature, flux and luminosity results compare quite well
with our global values.  The global fits show that {\it ASCA}
constrains both the thermal plasma temperature and heavy metal
abundance better than {\it ROSAT} because of {\it ASCA}'s larger
energy range and superior spectral resolution.  Likewise, {\it ROSAT}
determines the absorbing column and cooling rates better than {\it
ASCA}, because of its lower energy range.

\section{Radial Variation of the X-Ray Spectrum} \label{sec:annular}

\subsection{Deconvolved ASCA Spectra} \label{sec:annular_deconv}

\placefigure{fig:section}

We determined the radial variation of the X-ray spectrum by
accumulating {\it ASCA} spectra in 3 concentric annular regions
centered on the cD galaxy with radii of 0-1.5$^{\prime}$,
1.5-5$^{\prime}$, and 5-10$^{\prime}$.  In Figure~\ref{fig:section},
the annuli are shown superposed on a contour plot of the {\it ROSAT}
PSPC image. All of the annuli lie within the GIS FOV, while only the
inner two regions, $0-1.5^{\prime}$ and 1.5-5$^{\prime}$, fall fully
within the SIS FOV. Thus we exclude the SIS from analysis of the outer
region which only partially overlaps with the SIS detector. The annuli
were chosen such that each was larger than the {\it ASCA} XRT+GIS
PSF. We use methods developed by Markevitch (1996) to deconvolve the
emission spectrum of the gas from the energy dependent PSF of {\it
ASCA}.  Because of its higher spatial resolution, we use the {\it
ROSAT} PSPC image (Figure~\ref{fig:contour}) to constrain the relative
emission measure distribution between the regions.  However, because
of the possible contributions of a cooling flow or excess absorption
in the central region, we do not fix the relative emission measure of
this region from the {\it ROSAT} image, but allow it to vary.  Due to
the poor calibration of the {\it ASCA} telescope at low energies, we
restrict our analysis to energies in the range 1.5-10 keV, excluding
the gold edge from 2.0-2.5 keV.  In addition to statistical errors, we
include estimates of the systematic uncertainties of the background
(20$\%$, 1 $\sigma$), the PSF model's core (5$\%$) and wings (15$\%$),
and the effective area of {\it ASCA} (5$\%$), as suggested by
Markevitch (1996).  We also include the effect of the uncertainty in
the offset between the {\it ASCA} and {\it ROSAT} images.  The spectra
from each annulus and instrument were binned to insure that there were
at least 20 counts per energy bin.

As evidenced from Figure~\ref{fig:section}, there is a moderately
bright point source in the {\it ROSAT} PSPC image roughly
$11^{\prime}$ east of Abell 644. To ensure that this source is not
contaminating the cluster emission in the outer regions, given {\it
ASCA's} broad PSF, we made an energy redistribution matrix which
included this source as a separate region. For the source to truly be 
a contaminate, it should make up a significant percentage of the flux 
received from other regions.
Our analysis indicates that this source contributes
less than 1$\%$ of the flux in any of the regions used to accumulate the
cluster spectrum.

\placetable{tab:annular_asca}

The results of fitting the annular spectra 
are listed Table~\ref{tab:annular_asca}.
The values of $\chi^2$ given in the first row of the Table are for the
simultaneous fit to all of the annuli.
The {\it ASCA} temperatures are also plotted as a function of radius
in Figure~\ref{fig:temp}.
The minimum value for $\chi^2$ for all of the deconvolution fits
of {\it ASCA} spectra are unrealistically low because of the inclusion
of systematic errors (Markevitch 1996), which we have estimated 
pessimistically.

\placefigure{fig:temp}

For the Single Temperature Model, the temperature in the central
region is lower than that in the middle annulus, leading to a
non-monotonic radial temperature variation.  This low central
temperature could be due to the reduction of the mean temperature in
the central region by a cooling flow.  We also consider the
possibility that there is excess absorption toward the cooling flow.
The inclusion of excess absorption alone in the spectrum of the central
0-1.5$^{\prime}$ region improved the fits somewhat, but the excess
column is consistent with zero at the 90$\%$ confidence level.  The
best-fit value for the excess absorption of $\Delta N_{H} = 1.32
\times 10^{20}$ cm$^{-2}$ is again consistent with interpolation
errors of the galactic absorption value or from the additional
molecular component of the absorbing column.

Given the previous evidence from the X-ray surface brightness and spectra
for a cooling flow at the center of A644, we included a cooling
flow in the spectrum of the central 0-1.5$^{\prime}$ region in addition 
to the extra, intrinsic absorption.  As in \S~\ref{sec:global}, 
the gas was assumed to cool isobarically subject
to its own radiation from the ambient temperature in that region to a
very low temperature and had the same abundances as the
ambient gas. The results of the Cooling Flow Model fits are also
listed in Table~\ref{tab:annular_asca}.  The inclusion of the central
cooling flow increased the ambient gas temperature in the central
region, but had no significant effect on the temperatures in the outer
two annuli.  The values of $\chi^2$ were not reduced significantly from
those of the Excess Absorption Model.
The cooling rate was poorly determined.

\placefigure{fig:abun}

We used both fixed ($Z = 0.30 Z_{\odot}$) and variable abundance
models, and found that they are not statistically different. The
results of the variable abundance model are given in
Figure~\ref{fig:abun} and Table~\ref{tab:annular_asca}.  The middle
annulus has a lower abundance, but it is possible that this is an
artifact produced by the energy-dependent PSF of {\it ASCA}.  In any
case, the abundance is constant within the errors, even with the lower
abundance in the middle annulus.

\subsection{ROSAT PSPC Annular Spectra} \label{sec:annular_pspc}

\placetable{tab:annular_pspc}

The {\it ROSAT} PSPC spectra were accumulated in the same annuli as
were used for the {\it ASCA} spectra.  The results of fits to these
spectra are shown in Table~\ref{tab:annular_pspc}.  As noted in the
discussion of the global spectrum (\S~\ref{sec:global}), the PSPC
spectra can strongly constrain the absorption or the cooling rate, but
are relatively insensitive to the ambient gas temperature or the iron
abundance.  The central temperatures in the Single Temperature Model
or the model with Excess Absorption disagreed strongly with the {\it
ASCA} central temperature.  For the outer two annuli, the {\it ROSAT}
and {\it ASCA} temperatures were in reasonable agreement.  The iron
abundances, on the other hand, were rather high and almost
unconstrained.

Allowing for excess absorption didn't bring the {\it ROSAT} and {\it
ASCA} temperatures into agreement.  The excess absorption did not
increase toward the center of the cluster, as has been found in other
cooling flow clusters.  This suggests that this excess absorption
simply indicates that the Galactic absorption toward this cluster is
underestimated by the interpolation of the H~I measurements.

Finally, we included a cooling flow component in the spectra of the
two inner regions; the results appear in the rightmost columns of
Table~\ref{tab:annular_pspc}. This model improved the fit of innermost
region as well as the intrinsic absorption model did, but did not
significantly change the fit of the $1\farcm5-5\arcmin$ region.  The
temperatures, although very poorly constrained, are now consistent
with the {\it ASCA} values. We derive a cooling rate of 146 $(<380)$
$M_\odot$ yr$^{-1}$ within the inner $0\arcmin-1\farcm5$ region and 93
$(<392)$ $M_\odot$ yr$^{-1}$ for the $1\farcm5-5\arcmin$ region.

\section{Azimuthal Variations in the X-ray Spectrum} \label{sec:sections}

\placefigure{fig:temp_sect}

\placefigure{fig:abun_sect}

In order to search for azimuthal variations in the X-ray spectrum of
A644, we further divided the second and third annular regions into
four and two sectors, respectively.  The geometry of the sectors is
shown in Figure~\ref{fig:section}.  The spectra were simultaneously
fit to Single Temperature models, with the abundances allowed to vary
(fixing abundance at 0.30 yielded results which were not statistically
different from the variable abundance fits).  The results of the
variable abundance model are shown in Table~\ref{tab:sector}.  The
temperatures and abundances are also plotted in
Figures~\ref{fig:temp_sect} and \ref{fig:abun_sect}, respectively.  We
also did fits with central excess absorption and a central cooling
flow, but these did not affect the parameters in the outer sections.

\placetable{tab:sector}

Table~\ref{tab:sector} and Figure~\ref{fig:temp_sect} show that the
high temperature found in the second annulus in the previous analysis
(Table~\ref{tab:annular_asca} and Figure~\ref{fig:temp}) is the result
of a very hot region to the west of the cluster center.  The
temperatures are inconsistent with an isothermal distribution or
azimuthally symmetric distribution at $>99.9\%$ level.  However the
temperatures are roughly consistent with an isothermal distribution if
we exclude regions 3 and 5.  On the other hand, the abundances are
poorly determined, and there is no significant evidence for any
spatial variation.

There is no point source in the {\it ROSAT} PSPC or HRI images in the
high temperature western region, so it is unlikely that this spectral
component is due to a background AGN or some other sources.  We also
examined the {\it ASCA} GIS and SIS images of this region in the 7-10 keV
band, and found no evidence of a point source.  This test was done to
check whether the high temperature might be due to a strongly absorbed
or highly variable AGN, which might have been unobservable by {\it ROSAT}.
The hard X-ray emission which gives the western region of
the cluster its very high temperature appears to be extended in the
hard band GIS images.

\section{X-ray Surface Brightness Profile and Masses} \label{sec:mass}

The {\it ROSAT} PSPC counts were extracted in annular regions of
increasing size (1\arcmin\ - 5\arcmin) such that each region had an
adequate number of counts. Bright point sources were excluded from
these regions and X-ray background was subtracted from an annular
region where cluster emission was negligible ($\approx 34-44\arcmin$).
We used the results of the Cooling Flow Model fits to the global PSPC
(Table~\ref{tab:global}) spectrum to convert the surface brightness
into a physical flux. We also corrected the surface brightness for the
nominal Galactic absorbing column of $N^{gal}_{H} = 8.45 \times
10^{20}$ cm$^{-2}$ (Stark et al.\ 1992). The resulting X-ray surface
brightness is shown in Figure~\ref{fig:xsurf}.

\placefigure{fig:xsurf}

The {\it ROSAT} PSPC X-ray surface brightness was de-projected to
determine the X-ray emissivity as a function of radius on the
assumption that the cluster was spherically symmetric (see below).
The gas density $\rho$ was determined from the X-ray emissivity and
the {\it ASCA} temperature profile obtained from the annular regions
for the Single Temperature model (Figure~\ref{fig:temp}). Because the
{\it ASCA} temperatures were only determined in three radial bins, we
linearly interpolated the temperature to determine the values at all
radii.  The gas density was integrated over the interior volume to
give the gas mass as a function of radius, $M_{gas} (r)$. The
assumption of hydrostatic equilibrium allows the gravitational mass
$M_{tot}$ within a radius $r$ to be determined from
\begin{equation} \label{eq:hydrostatic}
M_{tot} (<r) = - \frac{r k T(r)}{\mu m_H G}
\left[ \frac{d \ln \rho (r)}{d \ln r} +
\frac{d \ln T(r)}{d \ln r} \right] \, ,
\end{equation}
where $T$ is the gas temperature, and $\mu$ is the mean particle mass
in terms of the mass of hydrogen, $m_H$.

Because of the enhanced X-ray surface brightness to the SSW
(Figure~\ref{fig:contour}) and the very high temperature in region 3
to the west (Figure~\ref{fig:temp}), it seems unlikely that the
cluster is well-relaxed, at least in this area.  On the other hand,
the northern side of the cluster appears more regular.  Thus, we
only use sectors 1, 4, 5, and 7 (see Figure~\ref{fig:section}) to 
determine the gas density in de-projection. We used this modified gas 
density to calculate the gas and gravitational masses.

\placefigure{fig:mass}

Figure~\ref{fig:mass} shows the accumulated gravitational mass (filled
squares) and gas mass (crosses) profiles.  The statistical errors for
these masses cannot be easily assigned since the errors in the
emissivities at different radii are correlated by the
de-projection. Thus errors were generated by a Monte Carlo technique
(Arnaud 1988; Irwin \& Sarazin 1995).  We randomly selected X-ray
count values for each radial bin from a Poisson distribution with a
mean value equal to the number of counts in the actual data.  We also
generated a simulated temperature profile from randomly selected
temperature values for each radial bin from a Gaussian distribution
with a mean value equal to the actual temperature derived in that
radial bin.  We use these two profiles to simulate electron density,
gas mass and gravitational mass profiles.  We ran this Monte Carlo
simulation 1000 times, sorted the values, and chose the 50th and 950th
values to define the 90$\%$ confidence region. Figure~\ref{fig:mass}
shows that the gas mass (the lower profile) accounts for 15$\%$ of the
gravitational mass (the upper profile) at a radius of 300 kpc,
increasing to 20$\%$ at 1.2 Mpc.

\section{Cooling Mass Deposition Rate} \label{sec:mdot}

The cooling radius, $r_{c}$, is defined as the radius at which the 
integrated cooling time is equal to the age of the cluster, for which 
we assume an age of 10$^{10}$ yr. For every radius for which the 
de-projection gave a value of the gas density, we calculated the 
integrated cooling time in the gas. The heavy element abundance is 
fixed at 30$\%$ of solar. From the derived density profile, we 
obtained a cooling radius $r_{c}$ = 261$^{+376}_{-64}$ kpc 

We determined the cooling rate profile using the technique outlined in
Arnaud (1988), which assumes a steady-state cooling flow.  Using the
de-projected X-ray emissivity, we determine the X-ray luminosity
output from each de-projected spherical shell. The luminosity from a
given shell is assumed to result from the combination of gas cooling
out of the flow in that shell, and gas flowing into that shell but not
cooling below X-ray emitting temperatures. The rate at which mass
drops out of the flow in shell $i$, $\Delta \dot M_i$, is taken to be
\begin{equation} \label{eq:mcool}
\Delta \dot M_i=\frac{\Delta L_i-\dot M_{i-1}(\Delta
H_i+\Delta \phi_i)}{H_i+f_i\Delta \phi_i} \, ,
\end{equation}
where $\Delta L_i$ is the change in bolometric luminosity across shell
$i$, $\dot M_{i-1}$ is the rate at which mass passes through shell $i$
but does not cool out of the flow, $H_i = 5kT(r)/2\mu m_H$ is the
enthalpy per unit mass in shell $i$, $\Delta H_i$ is the change in
$H_i$ across that shell, $\Delta \phi_i$ is the change in potential
across shell $i$, and $f_i$ is a geometrical factor to allow for the
fact that mass cools out of the flow in a volume-averaged way.  We
determine the gravitation potential assuming hydrostatic equilibrium
as discussed in \S~\ref{sec:mass}.  This gives
\begin{equation} \label{eq:hydro_potential}
\phi(r) - \phi(0) = - \int_{0}^{r} \frac{ k T(r)}{\mu m_H}
\left[ d \ln \rho (r) + d \ln T(r) \right] \, ,
\end{equation}
where $\phi(0)$ is the potential at the center of the cluster.

\placefigure{fig:mdot}
 
The total integrated cooling rate at a distance $r$ is simply 
the sum of the mass dropping out of the flow in all the shells at radii 
less than $r_n$:
\begin{equation} \label{eq:mdot}
\dot M(<r_n) = \sum_{i=1}^{n} \Delta \dot M(r_i) \, .
\end{equation}
Figure~\ref{fig:mdot} shows the resulting integrated cooling rate
profile, along with the 5 and 95 percentile values from the 1000 Monte
Carlo simulations.  As noted above, this technique assumes a
steady-state cooling flow, and thus may only apply out to $r_c = 261$
kpc.  The total cooling rate out to this radius is 393 $M_{\odot}
yr^{-1}$.  This is consistent but somewhat larger than the values
derived spectroscopically.

\section{Conclusions} \label{sec:conclusion}

The global X-ray spectrum of A644 is well fit by a model with a gas
temperature of $7.69^{+0.22}_{-0.20}$ keV, and an abundance of
$0.30\pm0.03$ $Z_{\odot}$.  The global spectrum suggests that there is
a cooling flow in the cluster with a cooling rate of
$214^{+100}_{-91}$ $M_{\odot}$ yr$^{-1}$.  The spectrum also suggests
an absorbing column which is slightly larger than the interpolated
Galactic H~I column, but this may be due to small scale variations in
the H~I column or to molecular material.  The excess absorption is not
concentrated to the cluster center, so it is probably not intrinsic to
the cooling flow or cluster.

We also determined the spatial variation of the temperature, iron
abundance, and cooling rate.  There is no significant evidence for any
spatial variation in the iron abundance.  We find evidence for a
region of extremely hot gas (10-25 keV) located 1.5-5$^{\prime}$ to
the west of the cluster center.  This region is perpendicular to the
enhanced emission in the {\it ROSAT} PSPC X-ray image. We suggest that
the cluster is undergoing or has recently undergone a merger.  The
combination of a moderate cooling flow and evidence for a merger make
this cluster an interesting case to test the disruption of cooling
flows by mergers.  Better spatial resolution data (e.g., with {\it
Chandra} or {\it XMM}) is needed to confirm the merger in A644 and to
determine its geometry.

We determined the X-ray surface brightness profile excluding the hot
region to the west and the extended region to the SSW.  We use
the X-ray surface brightness and temperature profiles to determine the
gas and gravitational masses as a function of radius.  The total
gravitating mass within 1.2 Mpc is $6.2\times10^{14} \, M_\odot$, of
which 20$\%$ is in the form of hot gas.  Since we have not included
the mass from individual galaxies, this result gives a lower limit on
the baryonic fraction within the cluster well above the upper limit
from cosmic nucleosynthesis of 0.06 for an $\Omega = 1$ universe
(e.g., Walker et al.\ 1991).  The baryonic fraction of A644 is
consistent with other clusters and groups (e.g., Allen \& Fabian 1994;
David et al.\ 1994).

\acknowledgements

This research has made use of data obtained through the High Energy
Astrophysics Science Archive Research Center Online Service, provided
by the NASA/Goddard Space Flight Center.
F. E. B. and C. L. S. were supported in part by NASA ASCA grants NAG 5-4516
and NAG 5-8390.
C. L. S. was also supported in part by NASA Astrophysical Theory Program
grant NAG 5-3057.
We would like to thank Maxim Markevitch for many helpful comments on
the analysis of {\it ASCA} data. F. E. B. would like to thank Jimmy
Irwin to his many helpful comments.

\clearpage

\figcaption[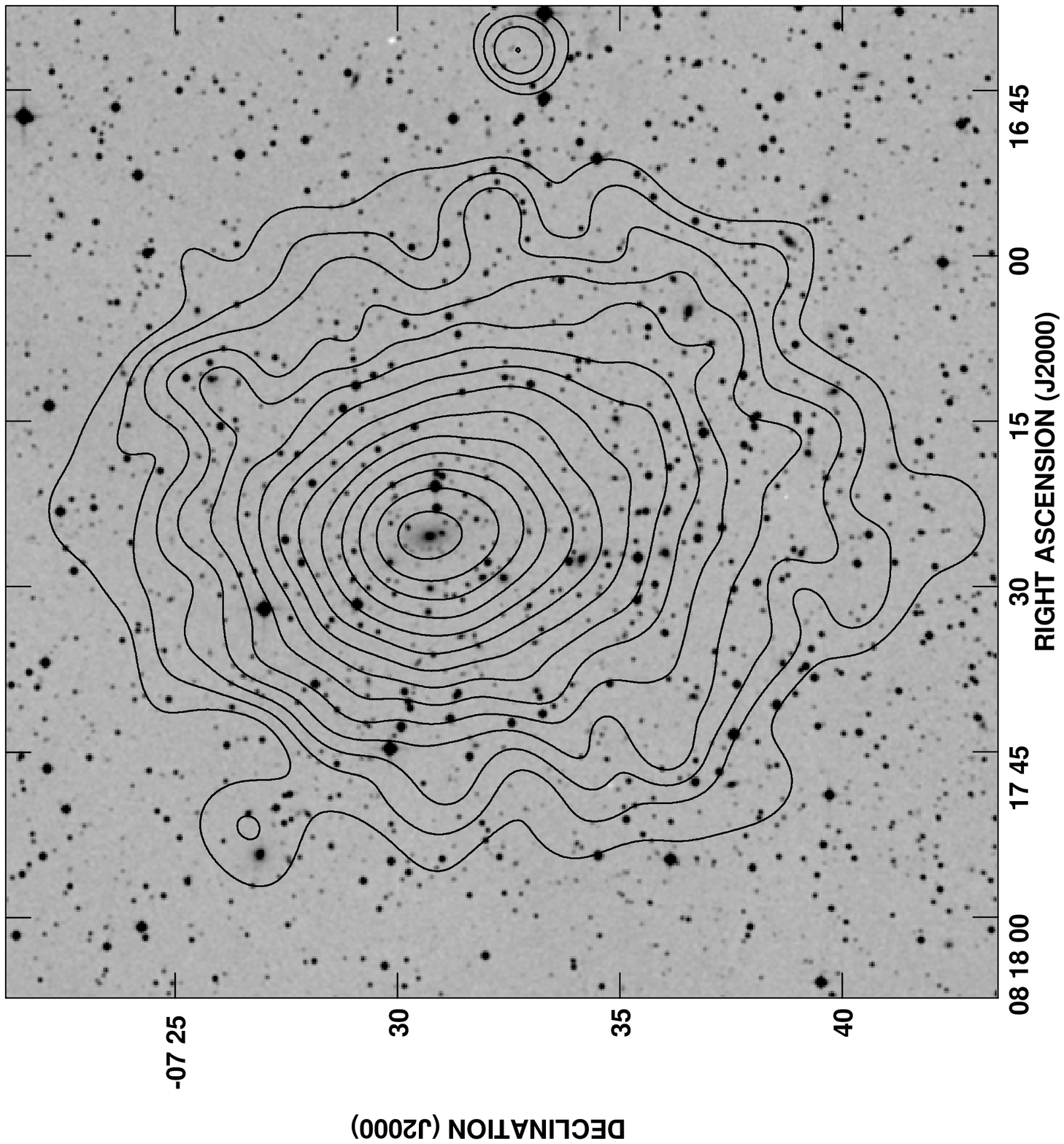]{
A contour plot of the {\it ROSAT} PSPC X-ray image in the 0.5--2.0 keV
band after adaptive kernel smoothing.  A minimum signal to noise ratio
of 5 per smoothing beam was used.  The contours are superposed on an
optical image from the Digital Sky Survey (Lasker et al.\ 1990).  The
lowest contour level is 0.4 counts/pixel ($7\farcs5$), with higher
contours being spaced logarithmically (by a factor of $2^{1/2}$).
Note the overall elliptical shape of the emission, which is elongated
in the same direction as the central cD galaxy, and the region of
enhanced X-ray emission about 5$^{\prime}$ to the south--southwest of
the cluster center.
\label{fig:contour}}

\figcaption[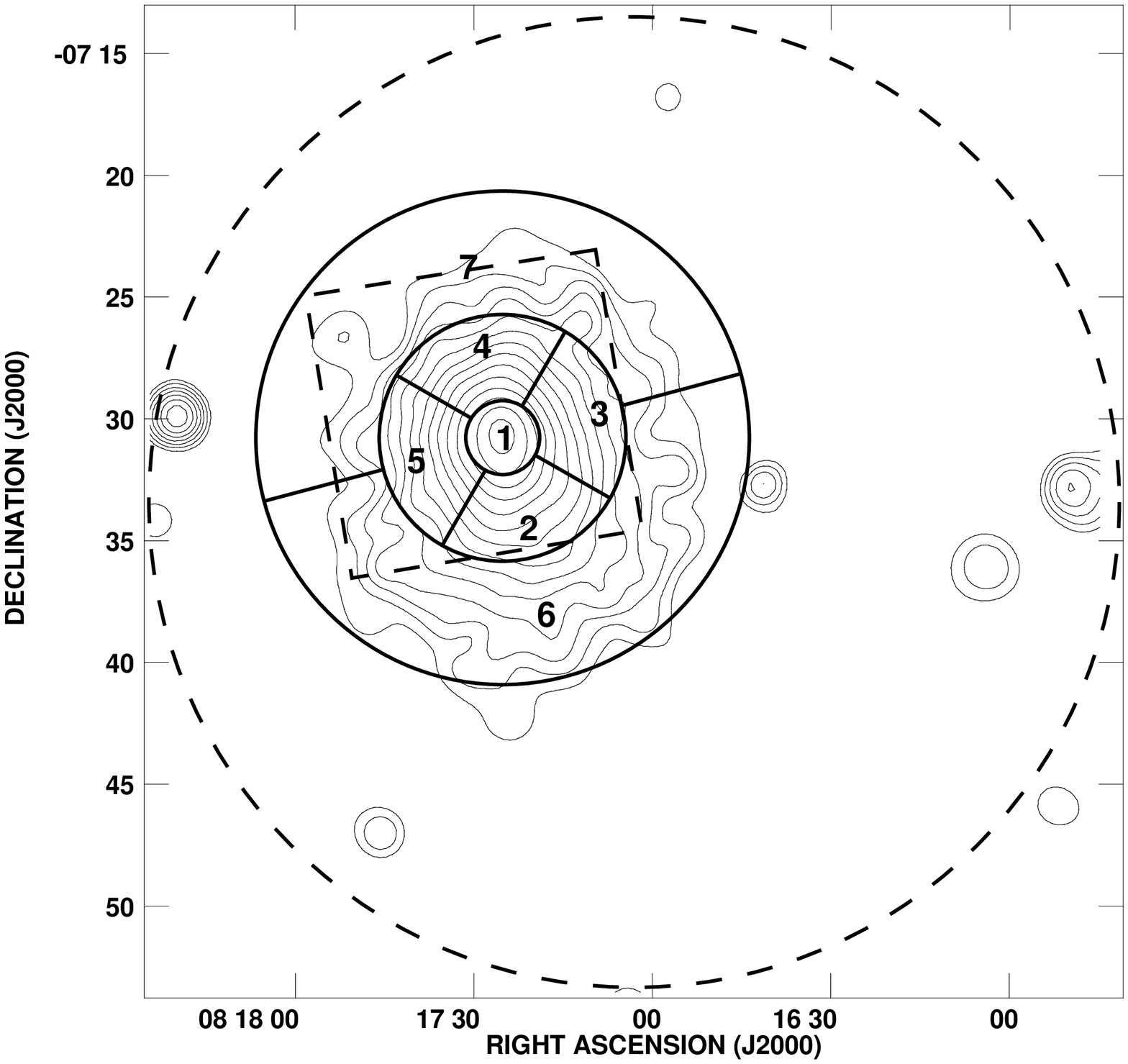]{
A contour plot of the the {\it ROSAT} PSPC X-ray image identical to
(Figure~\protect\ref{fig:contour}) is shown.  The dashed circle and
square represent the FOV of the GIS2 and SIS0 instruments,
respectively. Superposed are the annuli (0-1.5$^{\prime}$,
1.5-5$^{\prime}$, and 5-10$^{\prime}$) and sectors (1-7) used in the
spectral fitting to determine the radial and azimuthal variation of
the X-ray spectrum of A644.
\label{fig:section}}

\figcaption[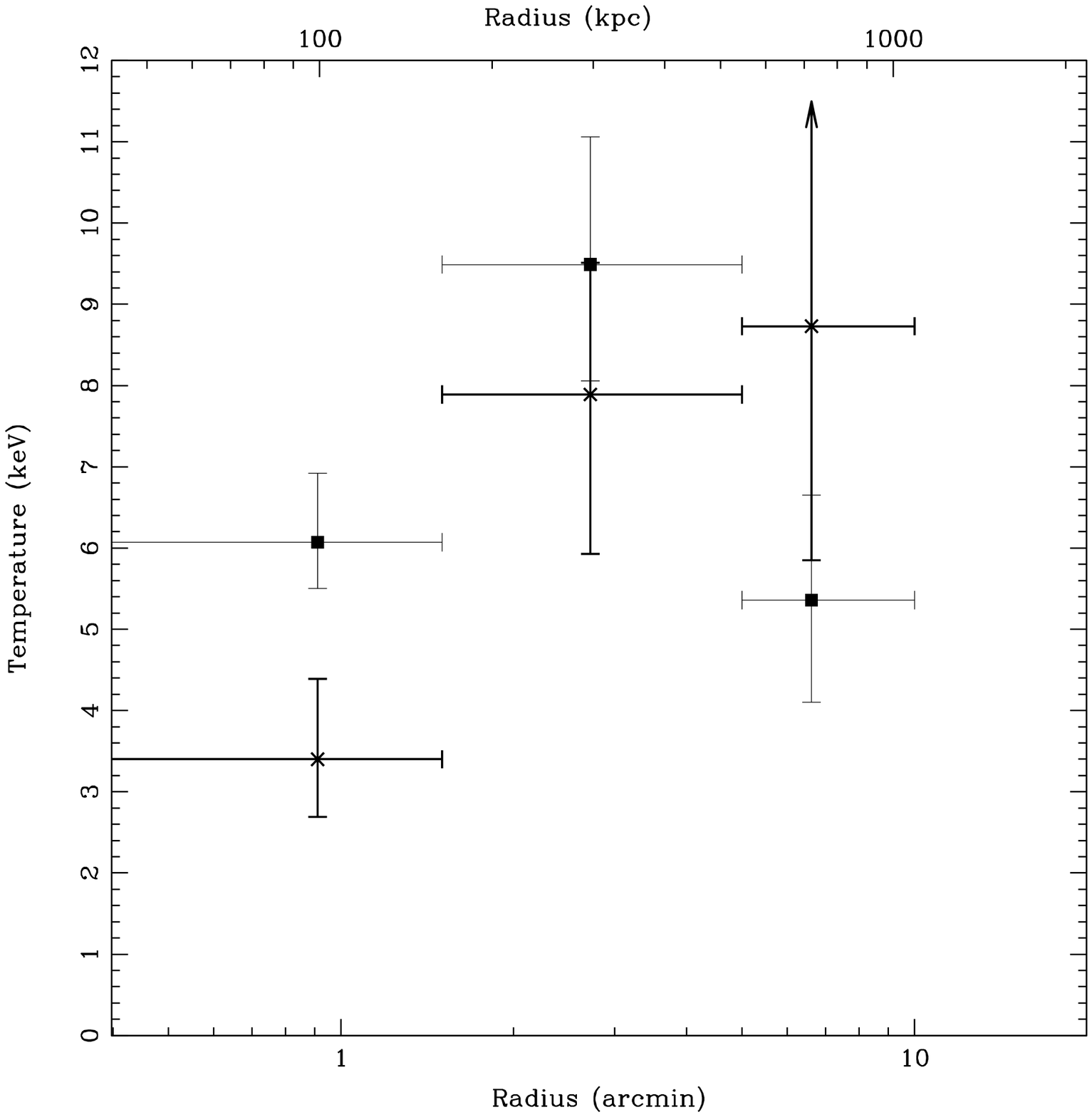]{
The radial temperature profile in A644 determined from single
temperature fits without excess absorption to {\it ASCA} spectra
(solid squares, light lines) and {\it ROSAT} PSPC spectra (diagonal
crosses, dark lines).  The horizontal error bars give the widths of
the annuli used.  The values are plotted at the emission weighted mean
radius derived from the {\it ROSAT} PSPC image.  The vertical error
bars represent the $90\%$ confidence regions.
\label{fig:temp}}

\figcaption[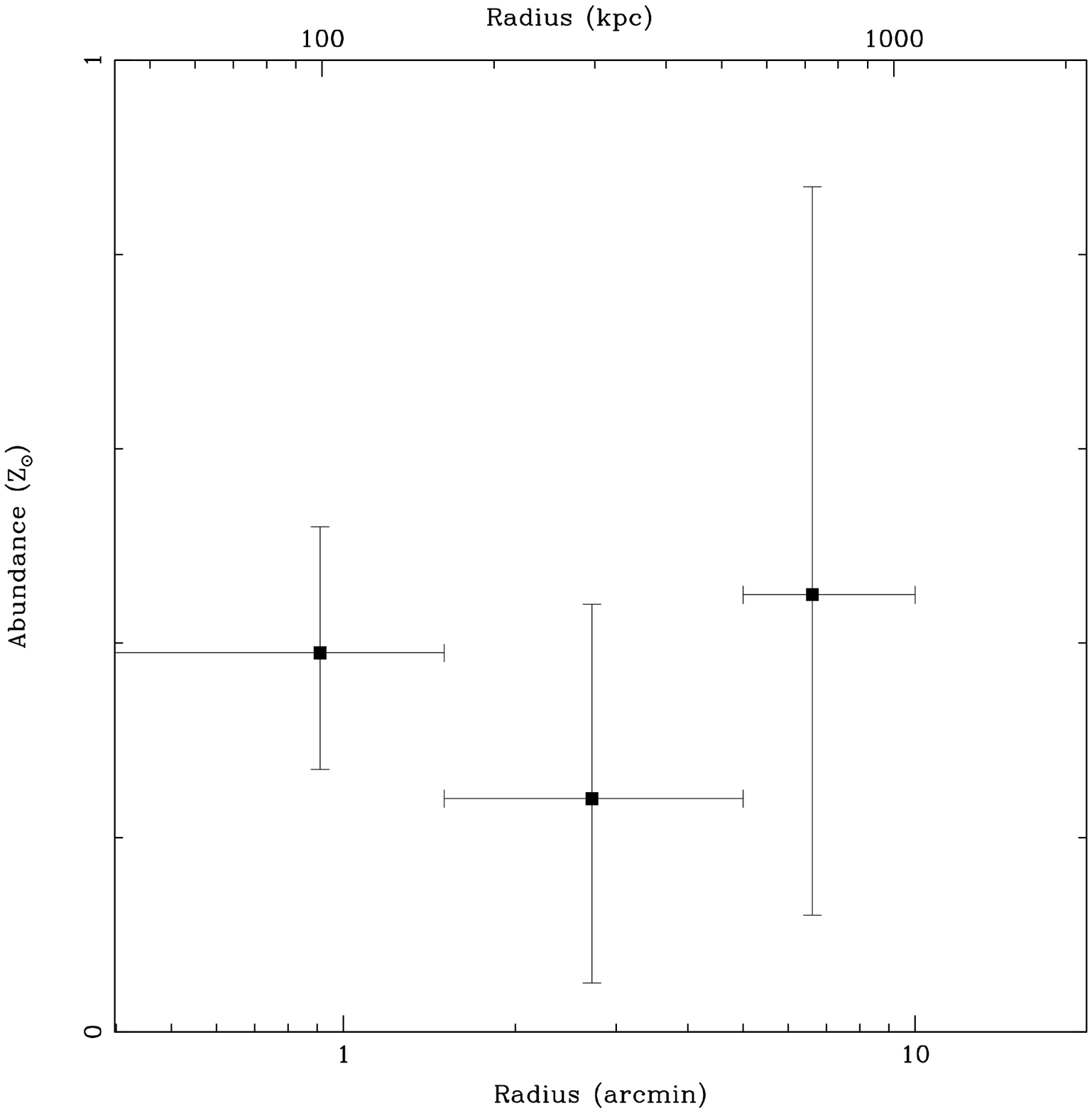]{
The radial profile of the heavy element (primarily iron) abundance in
A644 from the {\it ASCA} spectra.  The horizontal error bars give the
widths of the annuli used.  The values are plotted at the emission
weighted mean radius derived from the {\it ROSAT} PSPC image.  The
vertical error bars represent the $90\%$ confidence regions.
\label{fig:abun}}

\figcaption[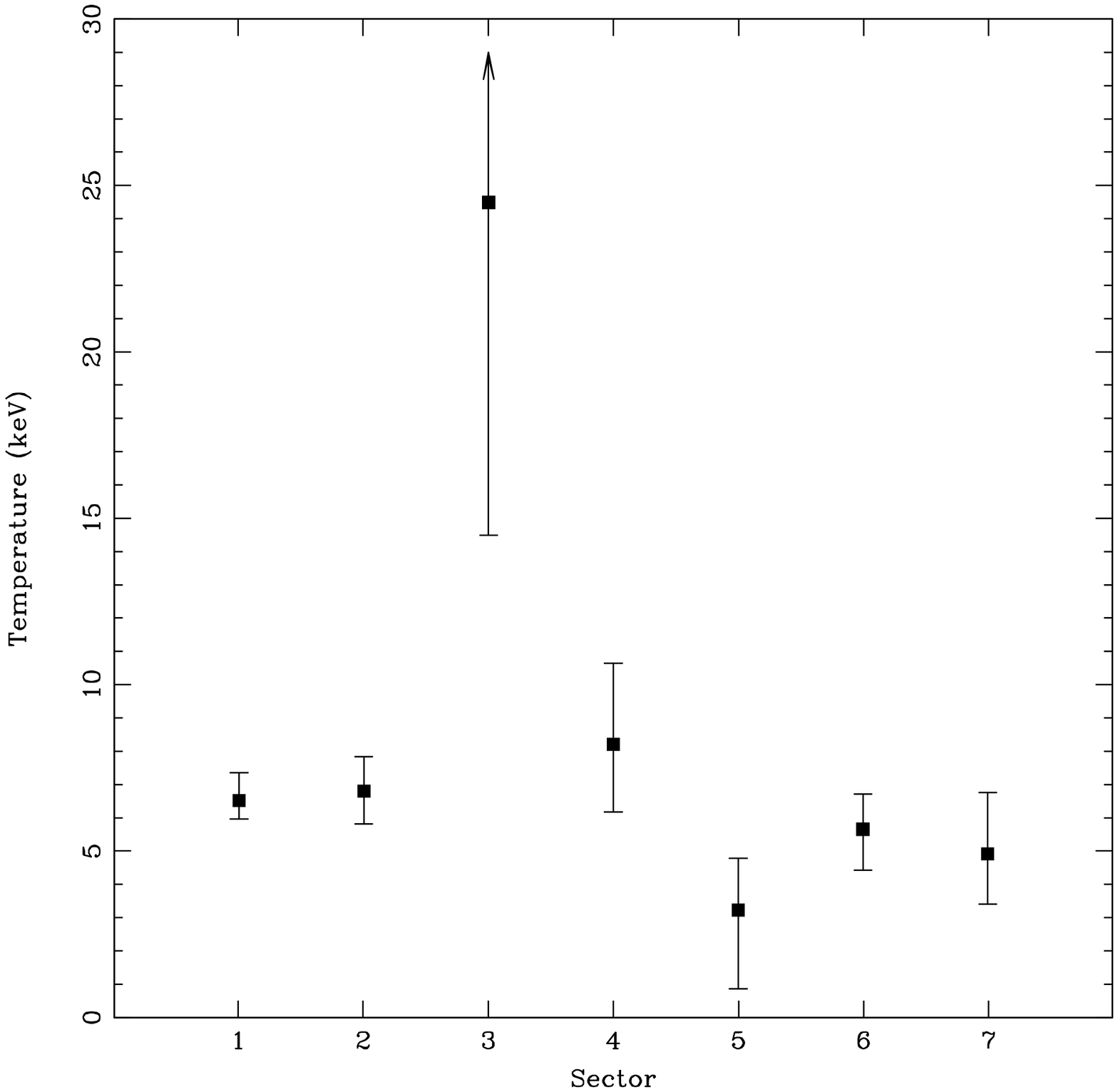]{
The temperatures determined from {\it ASCA} spectra for the sectors
shown in Figure~\protect\ref{fig:section}.  Error bars represent the
$90\%$ confidence regions.
\label{fig:temp_sect}}

\figcaption[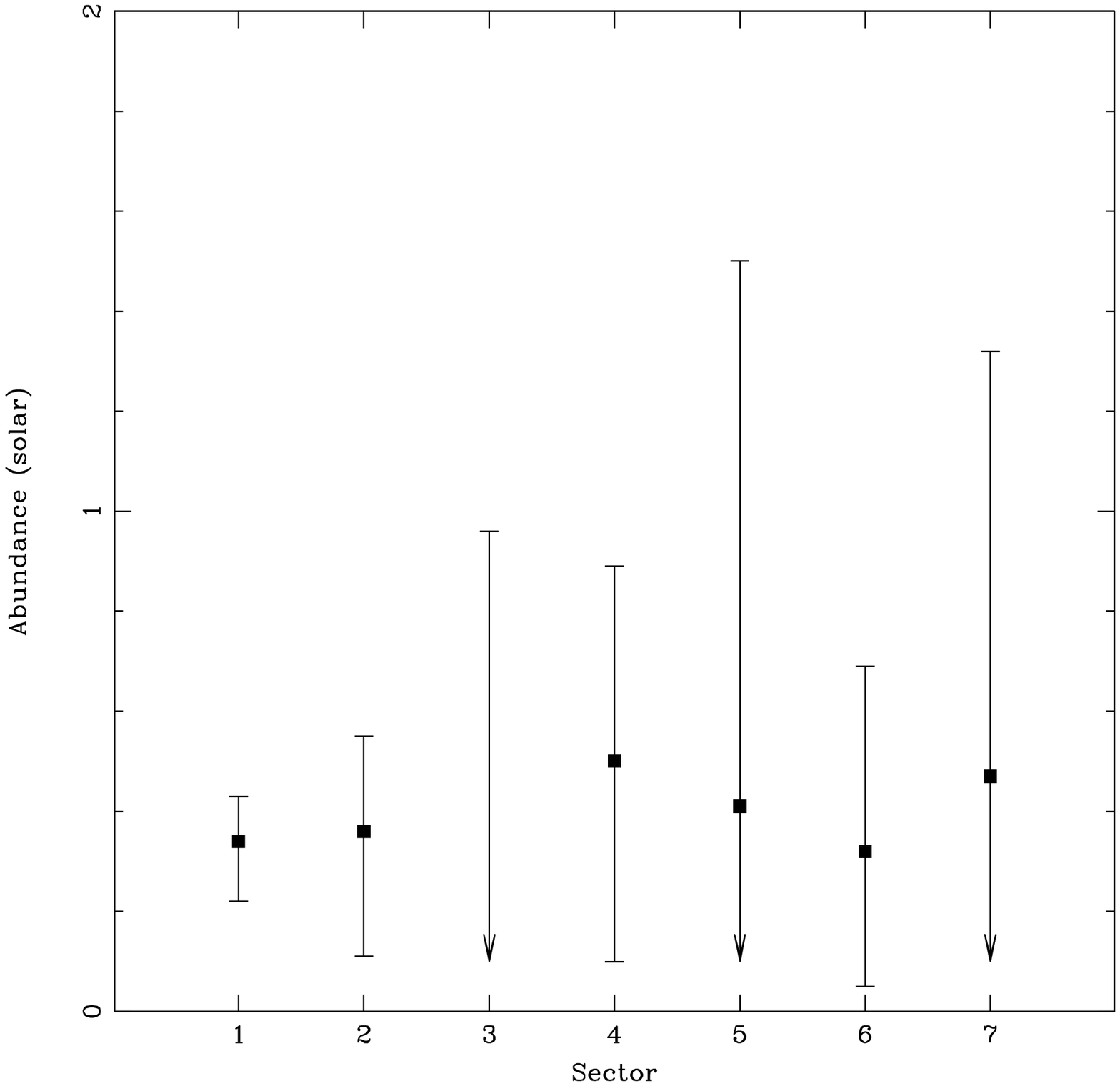]{
The abundances determined from {\it ASCA} spectra for the sectors
shown in Figure~\protect\ref{fig:section}.  Error bars represent the
$90\%$ confidence regions.
\label{fig:abun_sect}}

\figcaption[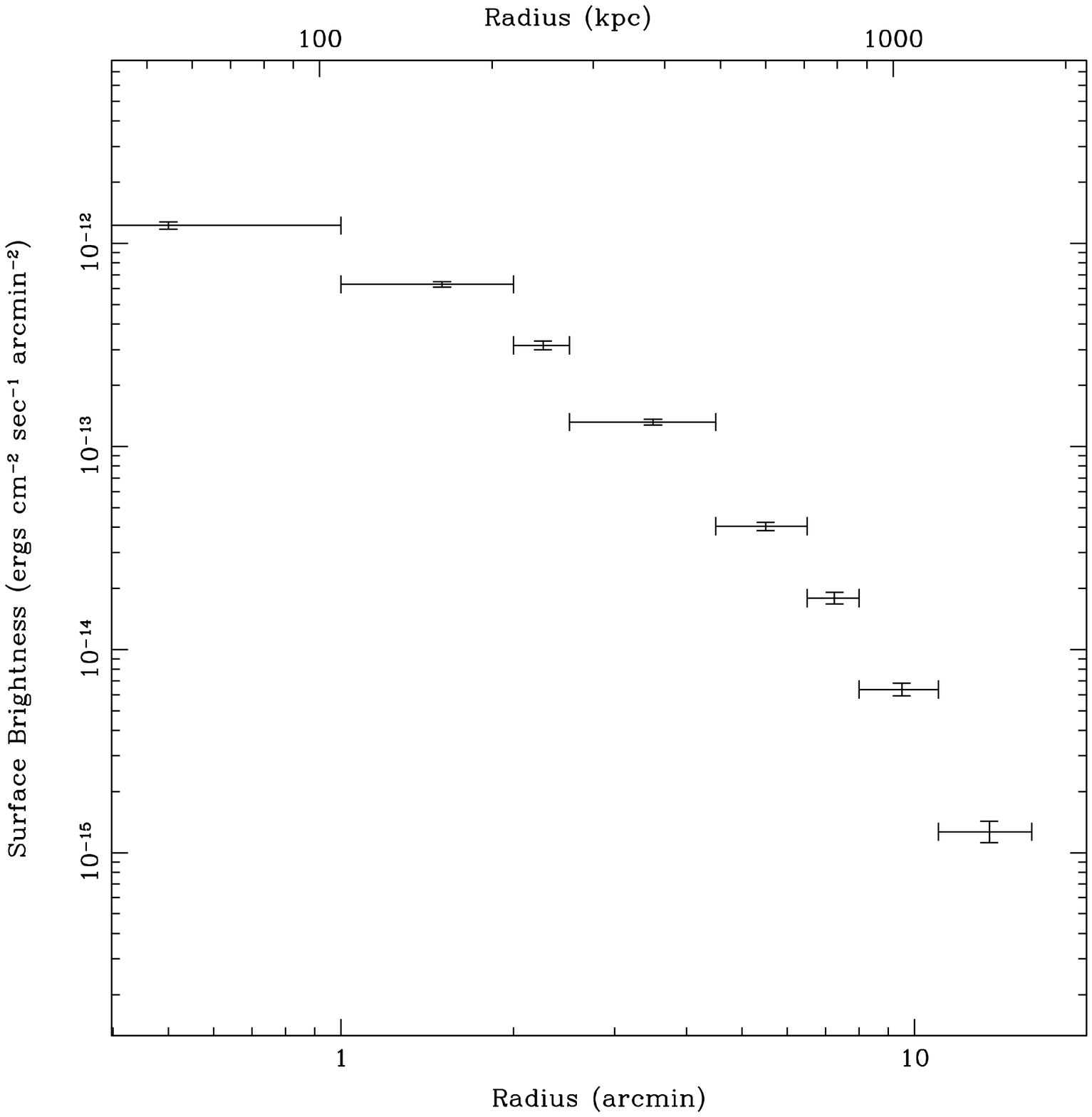]{
The {\it ROSAT} PSPC X-ray surface brightness of A644 in radial
annuli for the $0.5-2.02$ keV band with error bars representing the
$90\%$ confidence regions.
\label{fig:xsurf}}

\figcaption[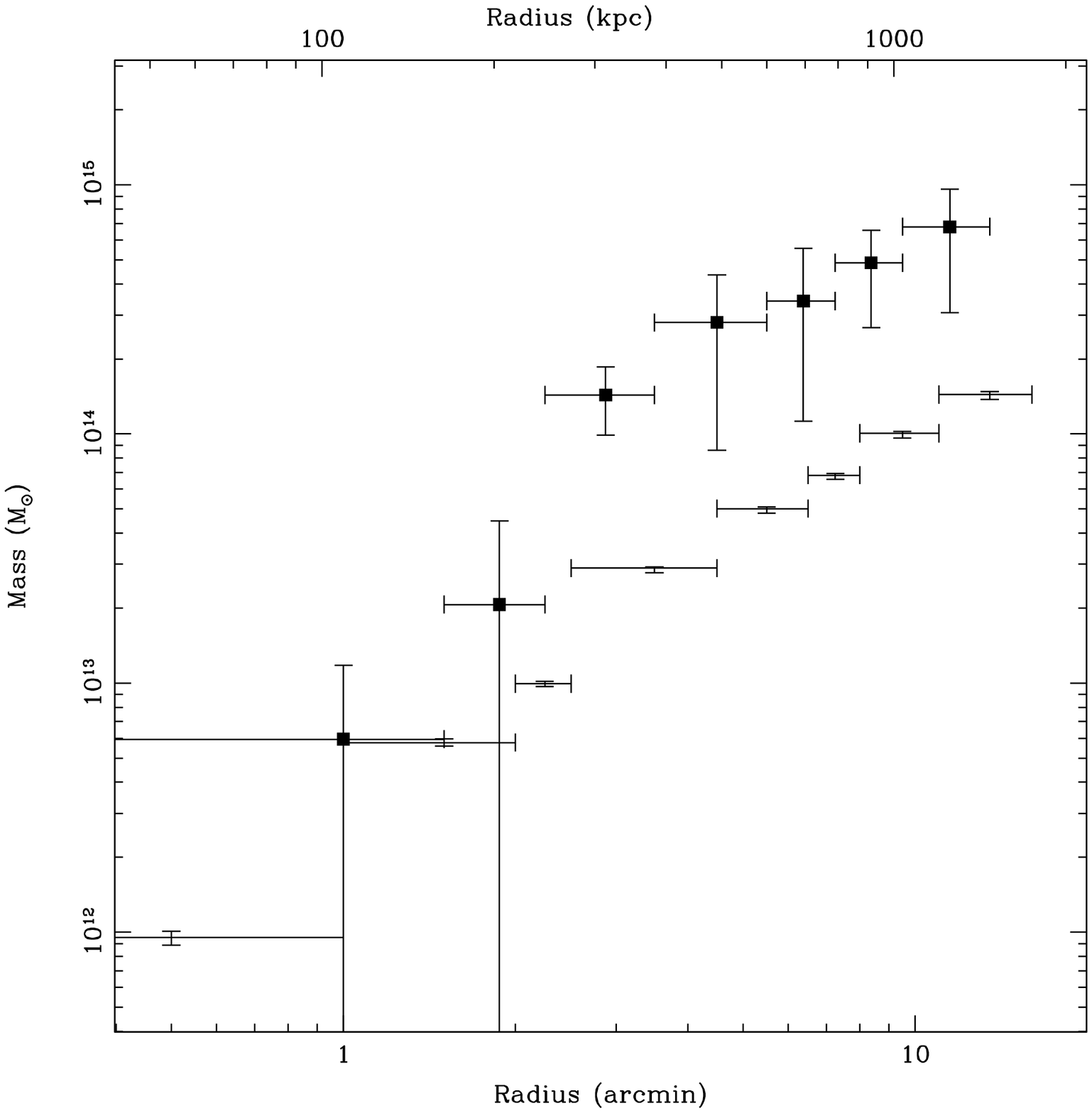]{
The values of the interior gravitational mass (filled squares, upper
points) and interior gas mass (no symbols, just error bars, lower
points) derived from the {\it ROSAT} PSPC radial surface brightness
profile and {\it ASCA} temperatures in annuli.  Error bars represent
the $90\%$ confidence regions from 1000 Monte Carlo simulations.
\label{fig:mass}}

\figcaption[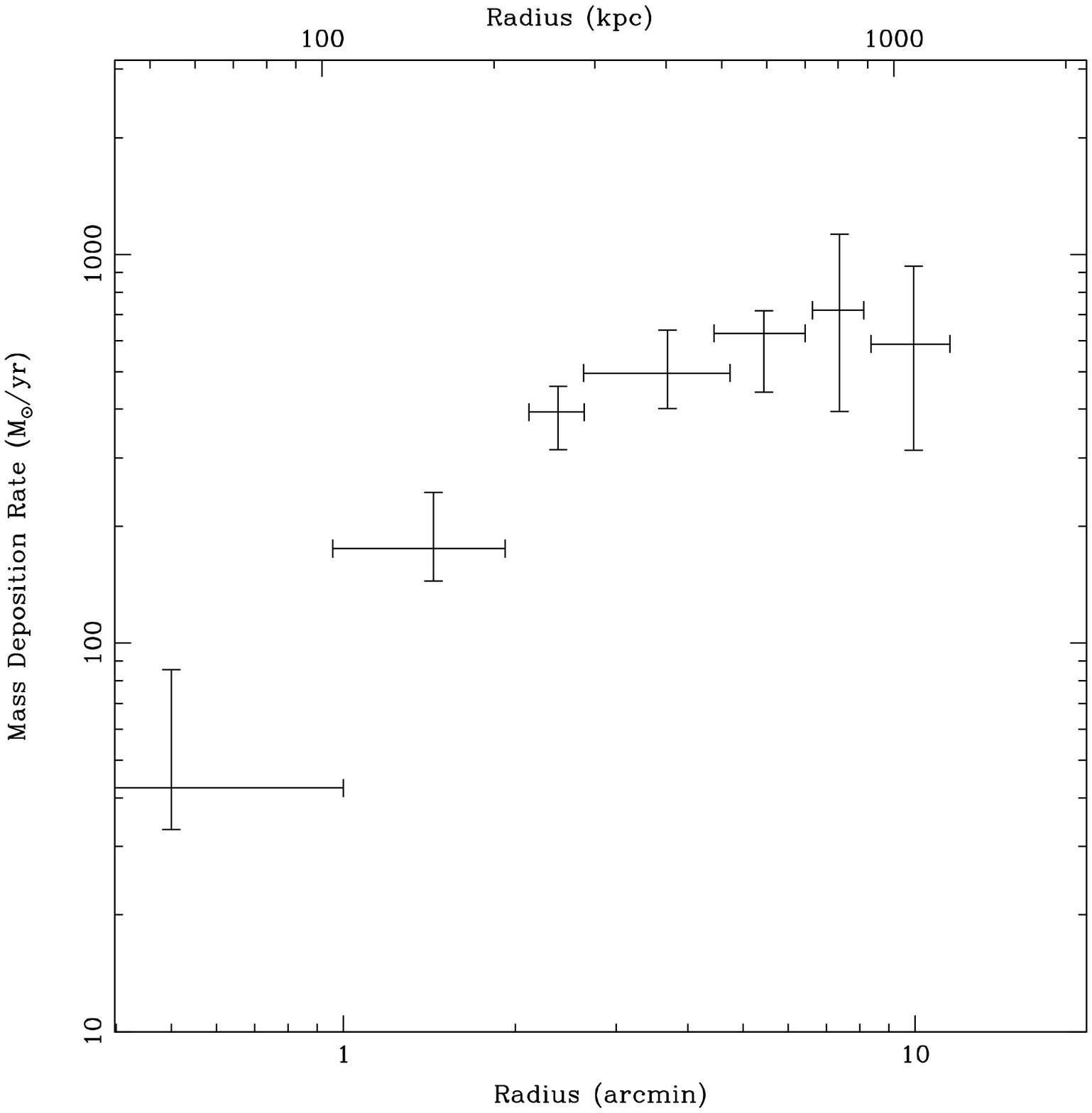]{
The integrated cooling rate profile for the cluster. Error bars represent
 the $90\%$ confidence regions from 1000 Monte Carlo simulations.
\label{fig:mdot}}

\clearpage

%
%
\begin{table}[htbp]
\dummytable\label{tab:global}
\end{table}
\clearpage
\begin{table}[htbp]
\dummytable\label{tab:annular_asca}
\end{table}
\clearpage
\begin{table}[htbp]
\dummytable\label{tab:annular_pspc}
\end{table}
\clearpage
\begin{table}[htbp]
\dummytable\label{tab:sector}
\end{table}
\clearpage

%
%

\begin{figure}[htbp]
\plotfiddle{fig1.eps}{5.0in}{-90}{80}{80}{-340}{370}
\end{figure}

\vspace{1.75truein}
\centerline{Figure~\ref{fig:contour}}

\clearpage

%
%

\begin{figure}[htbp]
\plotfiddle{fig2.eps}{5.0in}{0}{80}{80}{-260}{-120}
\end{figure}

 \vspace{1.75truein}
\centerline{Figure~\ref{fig:section}}

\clearpage

%
%

\begin{figure}[htbp]
\plotfiddle{fig3.eps}{5.0in}{0}{80}{80}{-250}{-220}
\end{figure}

\vspace{1.75truein}
\centerline{Figure~\ref{fig:temp}}

\clearpage

%
%

\begin{figure}[htbp]
\plotfiddle{fig4.eps}{5.0in}{0}{80}{80}{-250}{-220}
\end{figure}

\vspace{1.75truein}
\centerline{Figure~\ref{fig:abun}}

\clearpage
%
%

\begin{figure}[htbp]
\plotfiddle{fig5.eps}{5.0in}{0}{80}{80}{-250}{-220}
\end{figure}

\vspace{1.75truein}
\centerline{Figure~\ref{fig:temp_sect}}

\clearpage

%
%

\begin{figure}[htbp]
\plotfiddle{fig6.eps}{5.0in}{0}{80}{80}{-250}{-220}
\end{figure}

\vspace{1.75truein}
\centerline{Figure~\ref{fig:abun_sect}}

\clearpage

%
%
\begin{figure}[htbp]
\plotfiddle{fig7.eps}{5.0in}{0}{80}{80}{-250}{-220}
\end{figure}

\vspace{1.75truein}
\centerline{Figure~\ref{fig:xsurf}}

\clearpage

%
%
\begin{figure}[htbp]
\plotfiddle{fig8.eps}{5.0in}{0}{80}{80}{-250}{-220}
\end{figure}

\vspace{1.75truein}
\centerline{Figure~\ref{fig:mass}}

\clearpage
%
%
\begin{figure}[htbp]
\plotfiddle{fig9.eps}{5.0in}{0}{80}{80}{-250}{-220}
\end{figure}

\vspace{1.75truein}
\centerline{Figure~\ref{fig:mdot}}

\end{document}